\documentclass[twocolumn,preprintnumbers,amsmath,amssymb,superscriptaddress, altaffilletter, subeqn,prl, maxbibnames=99]{revtex4-2}
\usepackage{physics}
\usepackage{dcolumn}
\usepackage{bm}
\usepackage{graphicx}
\usepackage{amsmath}
\usepackage[dvipsnames]{xcolor}
\usepackage{float}
\usepackage{comment}
\usepackage{simpler-wick}

\begin{document}

\title{Exciton-Exciton Annihilation Mediated by Many-Body Coulomb and Phonon Interactions: An Ab Initio Study}
\author{Guy Vosco}
\affiliation{Department of Molecular Chemistry and Materials Science, Weizmann Institute of Science, Rehovot 7610001, Israel}
\author{Sivan Refaely-Abramson}
\affiliation{Department of Molecular Chemistry and Materials Science, Weizmann Institute of Science, Rehovot 7610001, Israel}

\begin{abstract}
Exciton-exciton annihilation (EEA), in which two excitons interact to generate high-energy excitations, is an important non-radiative channel in light-induced excited-state relaxation. When efficient, this process offers an alternative route to exciton emission, potentially allowing extended energetically excited particles' lifetime and coherence. These properties are significant in designing and understanding materials-based quantum devices, particularly for low-dimensional semiconductors. Here, we present a first-principles framework to compute EEA mechanisms and rates using many-body perturbation theory within the GW and Bethe-Salpeter Equation (GW-BSE) formalism. Our method explicitly accounts for Coulomb-driven and phonon-assisted exciton-exciton scattering by explicitly evaluating the interaction channels between the constituent electrons and holes composing the BSE excitons. We apply this framework to monolayer WSe$_2$ and explore the $A$, $B$ excitation manifolds, finding picosecond-scale annihilation between bright and dark states, cross valleys, and cross peak manifolds. These channels become allowed due to scattering into free electron-hole pairs across the Brillouin zone. Our results supply new insights into non-radiative exciton relaxation mechanisms in two-dimensional materials, providing a predictive and general tool for modeling these interactions in excitonic materials.
\end{abstract}

\maketitle

Excitons, bound electron-hole pairs \cite{cohen2016fundamentals}, serve as main energy carriers in emerging applications, such as material-based quantum information science and energy conversion and storage \cite{nielsen2010quantum,butler2013progress,wang2018colloquium,schaibley2016valleytronics}. The behavior of excitons, including their formation, interactions, and relaxation pathways, is strongly dictated by the structure and dimensionality of the host material. Low-dimensional semiconductors are a well-studied example, exhibiting nonuniform dielectric screening \cite{cudazzo2011dielectric,qiu2016screening,chernikov2014exciton} that leads to strongly bound excitons and long-lived excitonic states. These states decay in various pathways, including radiative recombination through photoluminescence (PL) as well as non-radiative processes involving scattering with other carriers, such as free electrons/holes, other excitons and exciton complexes, and phonons \cite{deotare20232d,moody2016exciton,chen2019ab,ginsberg2020spatially}. A key non-radiative process is exciton-exciton annihilation (EEA). In this Auger-like recombination mechanism, one exciton transfers its energy to another, leading to either free carrier generation or the formation of a highly excited bound exciton \cite{linardy2020harnessing,lin2021narrow,erkensten2021dark,steinhoff2021microscopic}. The efficiency of this pathway, along with radiative recombination and other scattering processes, plays a fundamental role in determining exciton lifetimes and the performance of optoelectronic and quantum devices.

Exciton-exciton annihilation (EEA) has been observed in a broad range of semiconducting materials, including quantum dots \cite{klimov2000quantization}, nanotubes \cite{wang2004observation}, and layered structures \cite{surrente2016onset,plaud2019exciton}. In monolayer transition metal dichalcogenides (TMDs), EEA is particularly significant due to strong excitonic effects and spin-valley coupling \cite{sun2014observation,kumar2014exciton,mouri2014nonlinear}. The lack of inversion symmetry and time-reversal symmetry constraints lead to unique excitonic selection rules, influencing their interactions and decay channels \cite{xiao2012coupled}. Prior theoretical studies have employed various approaches to study EEA in TMDs, including two-band k·p models \cite{konabe2014effect}, Monte Carlo simulations \cite{mouri2014nonlinear}, and effective Hamiltonian approaches \cite{chatterjee2019low}. A study by Steinhoff et al. \cite{steinhoff2021microscopic} on the involved many-body effects underscored the importance of accounting for the Fermionic nature of the exciton’s sub-particles for a complete understanding of EEA. However, a fully ab initio description incorporating both Coulomb- and phonon-driven scattering processes remains an open challenge.

Recent advances in computational methodologies and new theoretical derivations allow calculations of multi-particle excitations in low-dimensional semiconducting systems from a Green's function-based many-body perturbation theory approach within the GW and Bethe-Salpeter equation approximation (GW-BSE) \cite{deilmann2016three,biswas2023rydberg,torche2021biexcitons}. These methods give rise to new insights relating to spectral features observed in experiments and a theoretical association of these signatures with three- and four-particle exciton complexes. The interactions forming these exciton complexes are also the ones responsible for exciton-exciton scattering and can hence be used to study excited-state relaxation processes in layered semiconducting systems. To do this, a thorough and detailed investigation of the involved interactions and their influence on exciton relaxation is necessary.

In this work, we develop an ab initio framework to compute exciton-exciton annihilation, explicitly incorporating both Coulomb-mediated and phonon-assisted processes. Using a diagrammatic many-body perturbation theory approach, we derive the relevant coupling terms within an extended GW-BSE formalism. We apply this method to monolayer WSe$_2$, a typical TMD known for its strong excitonic interactions and multi-exciton phenomena. By analyzing the momentum-dependent interactions responsible for EEA, we provide a microscopic picture of the underlying scattering mechanisms and compute phonon-assisted EEA rates. We find the dominating annihilation upon light excitation include interactions between bright states to dark and inter-valley ones within the individual $A$ and $B$ peaks. Furthermore, we find a significant and unexpected contribution from interactions between these two peaks. We show that these arise due to wavefunction coupling and energy conservation, allowing the occupation of high-energy final states involving composite particles of non-interacting electron-hole pairs with various crystal momenta. 

\begin{figure}[H]
\includegraphics[width=1.0\linewidth]{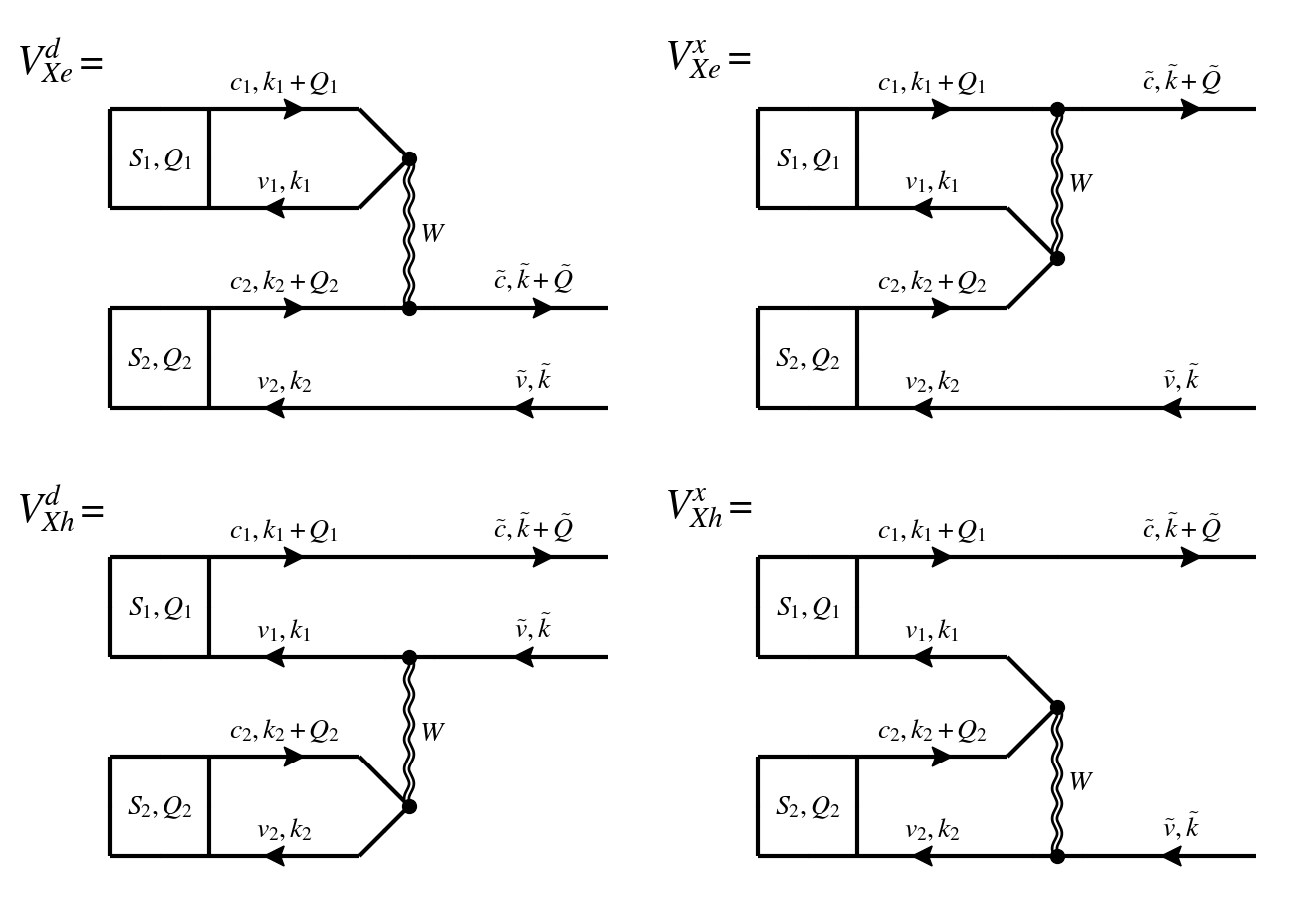}
\caption{Diagrammatic representation of the Coulomb-induced exciton-exciton annihilation interactions evaluated in this work. Solid lines with arrows going forward (and backward) denote the electrons (and holes), boxes represent the
incoming excitons, and wavy double lines correspond to the
screened Coulomb interaction. Time progresses from left to right. The initial state consists of two excitons with center of mass momenta of $Q_1$ and $Q_2$; the final
state consists of a free electron with momentum $\tilde{k}+\tilde{Q}$ and a hole with momentum $\tilde{k}$. The exchange of momenta $Q_1$ and $Q_2$ leads to a final Q-shift between the final electron hole pair of $\tilde{Q}=Q_1+Q_2$. Additional four interaction channels (not shown for brevity) that arise from a permutation of the two initial excitons are also considered.}
\label{Feynman diagrams}
\end{figure}

Figure \ref{Feynman diagrams} presents the interaction diagrams we consider for the EEA, for recombination of the electron-hole pairs composing the many-body GW-BSE excitons due to Coulomb interaction with an electron/hole of another exciton, while their pair hole/electron remaining unchanged. $S_i$ and $\mathbf{Q_i}$ represent the exciton states and their center-of-mass momentum, respectively; $v_i$,$c_i$ and $\mathbf{k_i}$ are the electron and hole states and the electron momentum; $W$ is the screened Coulomb interaction; $\tilde{c}$,$\tilde{v}$, are the final electron and hole states and $\tilde{\mathbf{k}}+\tilde{\mathbf{Q}}$ and $\tilde{\mathbf{k}}$ are the electron and hole momenta.  $V_{Xe}^d$ describes the recombination of an electron-hole pair originating from one of the excitons, induced by the interaction with an electron from an electron-hole pair originating from a second exciton. $V_{Xe}^x$ shows a similar interaction channel, but with the recombined pair consists of the electron from one exciton and the hole from the other. The superscripts $d$, $x$ refer to the 'direct-like' and 'exchange-like' nature of these interactions. $V_{Xh}^d$ and $V_{Xh}^x$ show the same recombination processes upon interaction with a hole instead of an electron. These terms, together with the permutations of the two initial excitons, comprise a total of eight diagrams.

As a starting point, we apply the G$_0$W$_0$ approximation on top of density-functional theory (DFT) to compute the quasiparticle wavefunctions and energies, respectively, using the Quantum Espresso \cite{giannozzi2020quantum,giannozzi2009quantum,giannozzi2017advanced} and BerkeleyGW \cite{hybertsen1986electron,rohlfing2000electron,deslippe2012berkeleygw} codes. Using these quasiparticle states for the electrons and the holes, we solve the BSE to obtain the excitonic eigenenergies and eigenfunctions. Within the GW-BSE framework, the spatial amplitude of an exciton $S$ with center-of-mass momentum $\mathbf{Q}$ is written as a linear combination of electron-hole pairs with $\mathbf{Q}$ momentum shift between the electron and the hole \cite{qiu2015nonanalyticity}, 
\begin{equation}
\ket{S,\mathbf{Q}} = \sum_{v,c,\mathbf{k}} A^{S,\mathbf{Q}}_{v,c,\mathbf{k}} a^\dagger_{c,\mathbf{k}+\mathbf{Q}} a_{v,\mathbf{k}} \ket{0}
\end{equation}
Where $A^{S,\mathbf{Q}}_{v,c,\mathbf{k}}$ are the expansion coefficients solving the BSE, $a^\dagger_{n,\mathbf{k}}$ and $a_{n,\mathbf{k}}$ are the creation and annihilation operators of particle in band $n$ with wavevector $\mathbf{k}$, and $\ket{0}$ is the ground state. We compute the EEA coupling terms:
\begin{equation} \begin{split}
V_{EEA}^{S_1;S_2}(v, c,\mathbf{k}) & = 
V^d_{Xe} - V^x_{Xe} - V^d_{Xh} + V^x_{Xh}  \\
\end{split} \end{equation}
where each coupling term consists of two diagrams- for initial states $\{S_1,S_2\}$ and $\{S_2,S_1\}$, such that the overall coupling contains the eight interaction terms, expressed in the summation:
\begin{equation} \begin{split}
V_{EEA}^{S_1;S_2}(v, c,\mathbf{k})
= \sum\limits_{\substack{i,j = 1,2 \\ i \neq j}} & \sum\limits_{\substack{v_i,c_i,\mathbf{k_i} \\ v_j,c_j,\mathbf{k_j}}} A^{S_i}_{v_i,c_i,\mathbf{k_i}} A^{S_j}_{v_j,c_j,\mathbf{k_j}} 
\\ \big(\big( &\hat{W}_{c_i,v_i;c,c_j} - \hat{W}_{c_i,c;v_i,c_j} \big) \delta_{v_j,v}
\\ + \big( & \hat{W}_{v,v_i;v_j,c_j} - \hat{W}_{c_j,v_i;v_j,v} \big) \delta_{c_i,c} \big)
\end{split} \end{equation}
Here, $\hat{W}$ is the screened Coulomb interaction matrix element, given by:
\begin{equation} \begin{split}
& \hat{W}_{n,m;n',m'} = \\ & \int{d^3r\ d^3r'} \psi^*_{n,\mathbf{k_n}}(r) \psi_{n', \mathbf{k_n}'}(r) W(r,r') \psi_{m,\mathbf{k_m}}(r') \psi^*_{m',\mathbf{k_m}'}(r').
\end{split} \end{equation}
We note that these matrix elements are computed to include interactions beyond the Tamm-Dancoff approximation (see SI for further details). 

Next, we compute the EEA scattering times:
\begin{equation}
    {\tau_{EEA}^{S_1,S_2}}^{-1} = \frac{2\pi}{\hbar N_k} \sum_{v,c,\mathbf{k}} |V_{EEA}^{S_1,S_2}(v,c,\mathbf{k})|^2 \rho(\Delta E)
\label{eq:rate}
\end{equation}
Where $N_k$ is the number of k-points in the computed numerical grid and $\rho(\Delta E)$ is the density of states, which in this case is represented as a delta function enforcing the conservation of energy and momentum,
\begin{equation}
    \Delta E = (\epsilon_{c,\mathbf{k}}-\epsilon_{v,\mathbf{k}+\mathbf{Q_1}+\mathbf{Q_2}}) - (\Omega_{S_1,\mathbf{Q_1}} + \Omega_{S_2,\mathbf{Q_2}})
\end{equation}
for $\epsilon_{n,\mathbf{k_n}}$ the electronic quasi-particle energies of a particle at band $n$ with momentum $\mathbf{k_n}$, and $\Omega_{S,\mathbf{Q}}$ the excitation energy of exciton $S$ with a center-of-mass momentum $\mathbf{Q}$. 

\begin{figure}[H]
\includegraphics[width=1.0\linewidth]{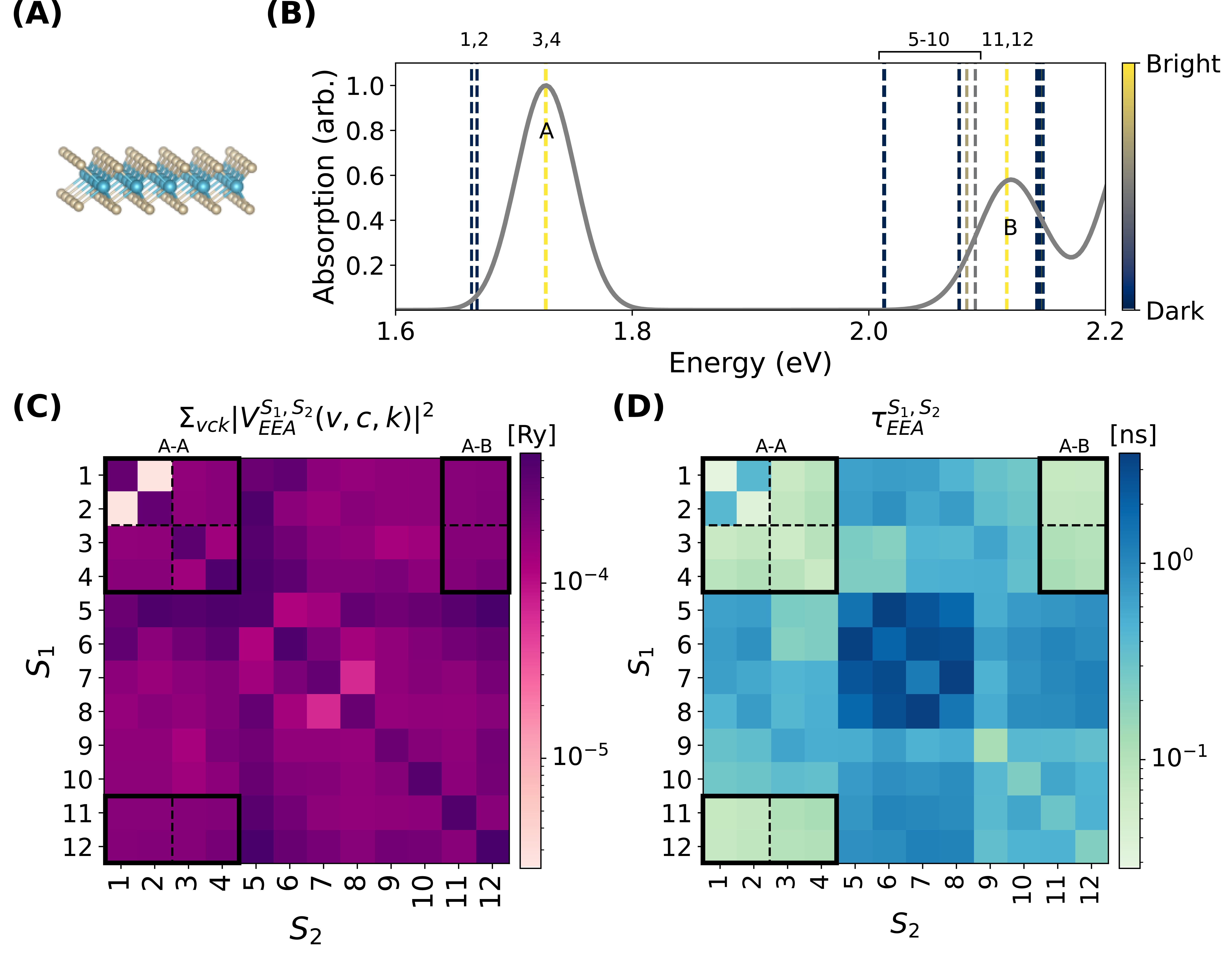}
\caption{(A) WSe$_2$ monolayer crystal structure, with Tungsten (W) atoms in blue and selenium (Se) atoms in yellow. (B) GW-BSE computed absorption spectrum, where each excitonic resonance is broadened by 40 meV. The excitation energies are shown in vertical lines, with colors indicating their relative oscillator strength, numbered by the exciton band indices up to the B peak region. (C) Calculated exciton-exciton annihilation coupling matrix elements, summed over all final states for a pair of initial excitons $S_1$ and $S_2$. (D) Computed exciton-exciton annihilation scattering times for pairs of initial excitons $S_1$ and $S_2$ within the examined energy range.}
\label{coul_results}
\end{figure}

The calculated WSe$_2$ monolayer crystal structure is shown in Figure~\ref{coul_results}(A). Panel (B) presents the calculated absorption spectrum, with the corresponding excitonic eigenenergies colored by their relative oscillator strength. The exciton indices correspond to their energy order in the BSE eigenvalue spectrum, with $1$ being the lowest-energy exciton. In WSe$_2$, exciotns 1,2 arise mainly from the transitions between holes in the highest valance band, $VB^+$, and electrons in the lowest conduction band, $CB^-$, around the two non-equivalents valleys $K$ and $\bar{K}$, forming spin-forbidden transitions that result in dark excitons. Excitons 3,4, primarily formed by transitions from holes in $VB^+$ and electrons in $CB^+$, are bright excitons that create the lowest-energy photoexcited resonance, the $A$ peak. Valley selectivity generates the following four dark states, excitons 5-8, and the $B$ peak is composed of excitons 9-12.

We focus on annihilation within and between the excitons constructing the $A$ and $B$ peaks, with excitation energies around 1.73 eV and 2.12 eV, respectively. We consider six conduction bands and four valence bands in the EEA process, which we find sufficient to ensure that energy conservation is not restricted. Figure~\ref{coul_results}(C) shows the resulting coupling matrix elements (in units of Rydberg) for the initial two excitons $S_1$ and $S_2$, summed over all final states. 
We find an orders-of-magnitude stronger coupling between excitons in which the electron-hole pairs are located in the same valley. The excitons of the opposite valleys are coupled rather weakly due to the small wavefunction overlap of these states. This manifests in larger values on the diagonal compared to the off-diagonals for the $A$, $B$ excitonic regions.
Figure~\ref {coul_results}(D) presents the corresponding computed scattering times, reflecting the interplay between the matrix element strength and the energy conservation between the initial excitonic states and the final free electron-hole pair, eq.~\ref{eq:rate}. 
Short scattering times are found for the $A$ peaks excitons, including the lower dark states excitons, while annihilation involving higher-energy dark excitons is suppressed. Additionally, our results predict short scattering times for the interaction between the $A$-peak excitons and $B$-peak excitons, suggesting a significant interaction channel between these states. 

\begin{figure}[]
\includegraphics[width=1.0\linewidth]{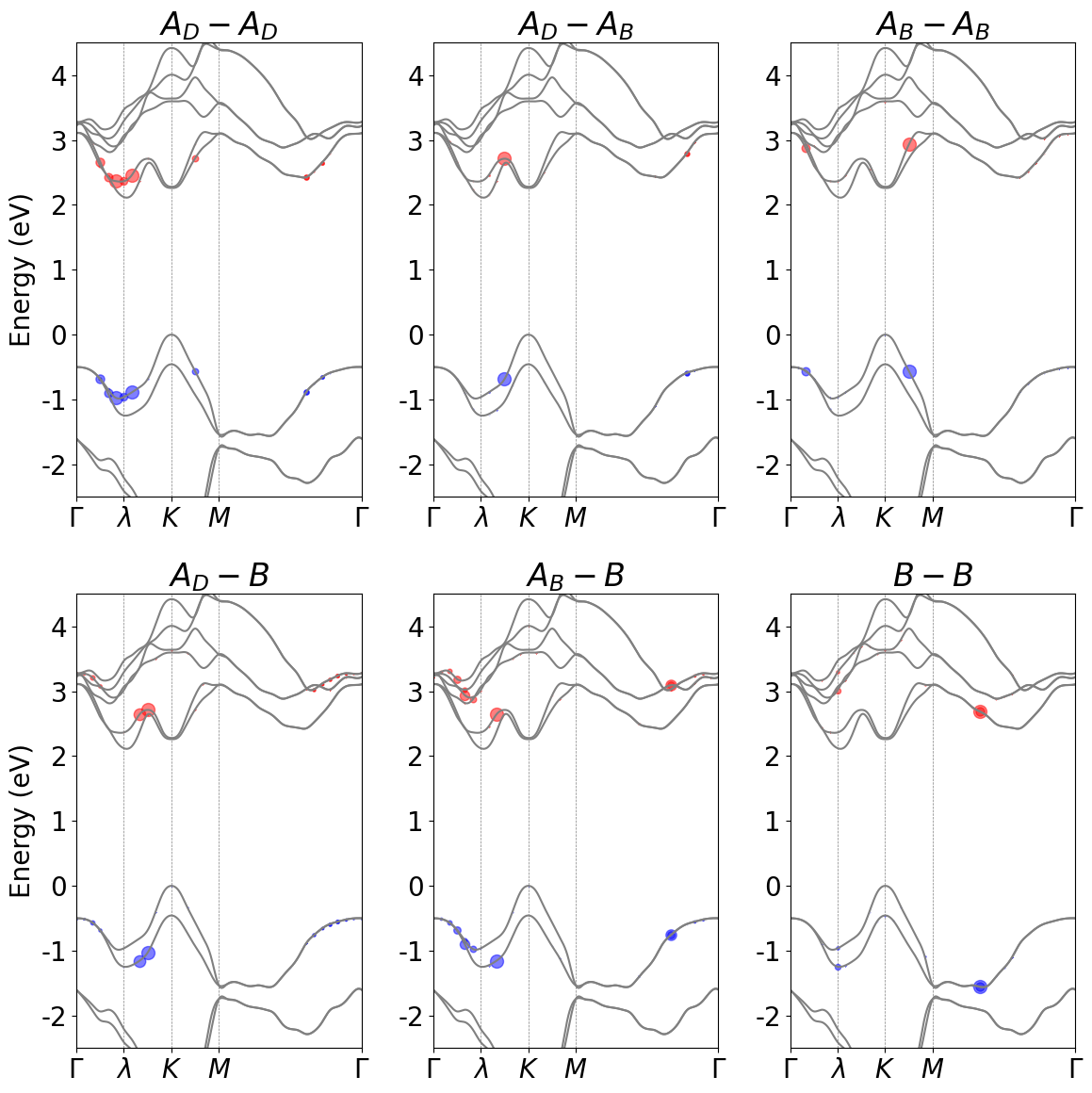}
\caption{
Final-state projections of the calculated exciton–exciton annihilation processes. Each panel shows the dominant bands contributing to the resulting free electron–hole pairs for a given initial exciton pair, plotted along high-symmetry paths in the Brillouin zone. Top row: annihilation of (left) two dark $A$ excitons ($A_D$), (middle) a dark and a bright $A$ exciton ($A_B$), and (right) two bright $A$ excitons. Bottom row: annihilation of $B$ peak excitons with (left) dark $A$, (middle) bright $A$, and (right) $B$ excitons. The contributions are normalized per interaction channel.
}
\label{EEA_BS}
\end{figure}

Figure~\ref{EEA_BS} visualizes the dominant final electron–hole band contributions resulting from representative initial exciton combinations along high-symmetry paths in the band structure. In all the examined scattering processes within this energy regime, the resulting final states are neither located at the K-valleys nor spin-distinct. Hence, the degenerate exciton pairs at $K$ and $\bar{K}$, such as excitons 1 and 2 or 3 and 4, are coupled to similar final states, allowing us to represent these excitonic pairs in the analysis. While these contributions are presented for a selected high-symmetry path, they serve as a guide to understand the overall trend (a more detailed analysis of the final states is shown in the SI).
The annihilation between the two lowest dark states, $A_D-A_D$, predominantly couples to electron–hole pairs in the two lowest conduction bands $CB^\pm$ and the top valence band $VB^+$ near the K valleys. However, because these transitions are energetically forbidden at $K$/$\bar{K}$, the final-state momenta shift toward the $\Lambda$ points, where energy–momentum conservation can be satisfied, as shown in the top left panel of Figure \ref{EEA_BS}.
The annihilation of a dark and a bright $A$ exciton, $A_D-A_B$, yields final states primarily in the second-lowest conduction band $CB^+$ and the top valence band $VB^-$ near the $K-\Lambda$ path (top center panel). The annihilation of two bright A excitons, $A_B-A_B$, results in final states located between $K$ and $M$, involving the conduction band $CB^-$ and valence band $VB^-$ (top right panel).
The second row in Figure \ref{EEA_BS} shows the final states of annihilation involving $B$ peak excitons. Interactions between $A$ and $B$ peak excitons yield final states near the $K-\Lambda$ region, while $B–B$ annihilation produces final states located along the $M-\Gamma$ path. 

Our calculations reveal final-state contributions involving high conduction bands such as $CB+2^\pm$. These should not necessarily be identified with scattering to high-lying excitons, as was previously suggested as a possible EEA outcome in WSe$_2$ \cite{lin2021narrow} for states associated with transitions near the $K$/$\bar{K}$ valleys, where strong optical activity and bound-state character are emphasized. Instead, the final states we find involving $CB+2^\pm$ appear at various momenta along the $M-\Gamma$ and $K-\Lambda$ paths, suggesting that these transitions are driven by momentum conservation and the many-body nature of the excitons, rather than forming tightly bound excitons. Moreover, such high-energy final states are observed only in annihilation processes involving $B$ peak excitons, with either $B–B$ or $B–A$ combinations. In contrast, annihilation between $A$ peak excitons, including both bright and dark states, predominantly yields final carriers in the lowest conduction and valence bands near the $K$ valleys. These characteristics are more consistent with the generation of dissociated electron–hole pairs than with the formation of bound high-lying states. Therefore, while our results do not rule out the possibility of high-energy excitonic final states under certain conditions, they suggest that, in the case of WSe$_2$, EEA more commonly results in unbound carrier generation, particularly for $A$ exciton–dominated processes.

Finally, we examine temperature effects on the EEA by deriving interaction terms for a combined process of phonon-assisted EEA within the framework of second-order perturbation theory. We treat the phonon-scattered excitons as an intermediate state that subsequently undergoes the annihilation process. Consequently, the effective phonon-assisted EEA matrix element is obtained by summing over all intermediate states, weighted by their energy differences. Within this framework, each exciton can gain momentum through phonon scattering before undergoing annihilation, thereby modifying the selection rules and introducing momentum shifts between the electron-hole pairs, opening up additional annihilation pathways. The interaction terms are given by
\begin{equation} \begin{split}
    V_{phEEA}^{S_1,S_2}(\nu,\mathbf{q};v,&c,\mathbf{k}) = \\
    \sum_{\tilde{S}} & \frac{G_{S_1,\tilde{S},\nu}(0,\mathbf{q}) V_{EEA}^{\tilde{S},\mathbf{q};S_2,0}(v,c,\mathbf{k})}{(\Omega_{S_1,0}-\Omega_{\tilde{S},\mathbf{q}} \pm \hbar\omega_{\nu,q}+i\eta)} \\
    + & \frac{G_{S_2,\tilde{S},\nu}(0,\mathbf{q}) V_{EEA}^{S_1,0;\tilde{S},\mathbf{q}}(v,c,\mathbf{k})}{(\Omega_{S_2,0}-\Omega_{\tilde{S},\mathbf{q}} \pm \hbar\omega_{\nu,q}+i\eta)}
\end{split} \end{equation}
where $\omega_{\nu,\mathbf{q}}$ are the phonon frequencies and $G_{S_i,\tilde{S},\nu}(0,\mathbf{q})$ are the exciton phonon coupling matrix elements between exciton $\ket{S_i,0}$ and exciton $\ket{\tilde{S},\mathbf{q}}$ scattered by phonon $\nu$, calculated using density functional perturbation theory \cite{chen2020exciton,antonius2022theory}. To compute the phonon-assisted EEA rates, we consider both phonon absorption and emission for each exciton involved in the interaction. The density of states of this process is
%\begin{equation}\begin{split}
%    \rho(\Delta &E;\nu,\mathbf{q}) = (N_{\nu,\mathbf{q}} + \frac{1}{2} \pm \frac{1}{2}) \delta(\Delta E \pm \hbar\omega_{\nu,\mathbf{q}})
%\end{split}\end{equation}
$\rho(\Delta E;\nu,\mathbf{q}) = (N_{\nu,\mathbf{q}} + \frac{1}{2} \pm \frac{1}{2}) \delta(\Delta E \pm \hbar\omega_{\nu,\mathbf{q}})$, with $N_{\nu,\mathbf{q}}$ the phonon Bose-Einstein occupation function at temperature $T$. 

\begin{figure}[]
\includegraphics[width=1.0\linewidth]{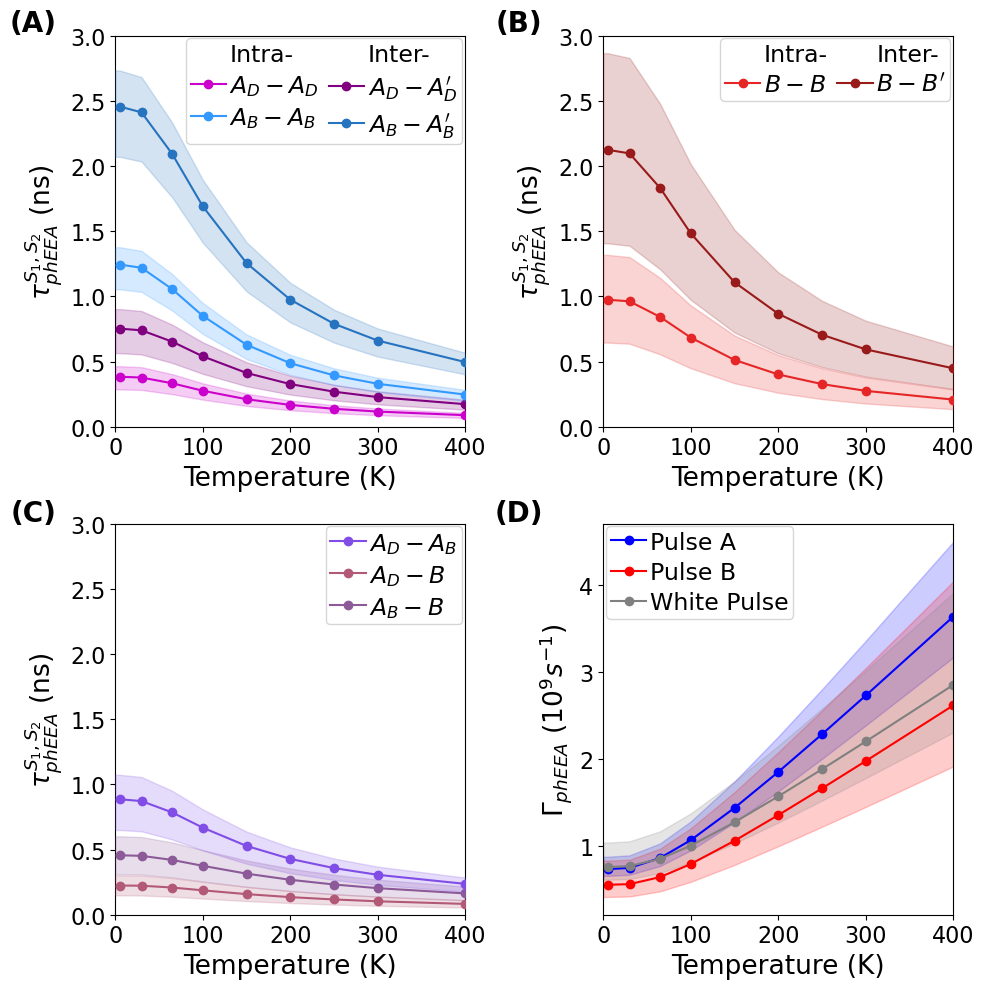}
\caption{(A) Phonon-assisted exciton-exciton annihilation scattering times for $A$-peak excitons as a function of temperature. The intra-valley interactions between dark ($A_D-A_D$) and bright ($A_B-A_B$) are compared to their respective inter-valley interactions ($A_D-A_D'$ and $A_B-A_B'$). The shaded regions represent the uncertainty arising from variation in the damping parameter $\eta$ between 10 meV to 20 meV. 
(B) The scattering time for the interactions of the $B$-peak excitons for the intra-valley ($B-B$) and inter-valley ($B-B'$) interactions.
(C) The scattering times of annihilation involving the combinations of dark and bright $A$-peak ($A_D$ and $A_B$) and B-peak excitons.
(D) Annihilation rate of the photoexcited populations, generated by setting Gaussian light pulses around the $A$-peak and the $B$-peak, and a broadband excitation (white pulse).}
\label{ph_results}
\end{figure}

The phonon-assisted EEA scattering times, presented in Figure~\ref{ph_results}(A-B), show that phonon scattering enables excitons in opposite valleys to undergo exciton-exciton annihilation more efficiently. At low temperatures, inter-valley processes are suppressed, while at higher temperatures, the increased phonon population allows for intermediate scattering states that activate these otherwise inaccessible pathways. As a result, inter-valley scattering times decrease and become comparable to those of the intra-valley processes. This temperature dependence reflects the role of phonons in relaxing momentum and spin constraints, effectively enabling annihilation processes that are otherwise forbidden.
Figure~\ref{ph_results}(C) shows interactions between excitons from the $A$ and $B$ peaks and their combinations. The annihilation processes involving the dark $A$ excitons ($A_D$) decrease the scattering times compared to bright ones, suggesting that these states contribute significantly to the overall annihilation rate as a phonon-assisted pathway. We focused on single-phonon processes, which do not allow the dark-dark annihilation pathway for photoexcited excitons, although it is generally possible to generalize our approach to include two-phonon-assisted processes as well. 

Finally, we connect the computed scattering times to phonon-induced decay rates. We set Gaussian-pulse light excitations, centered at 1.73 eV (\textit{Pulse $A$}) and 2.12 eV (\textit{Pulse $B$}) with a 25 meV full width at half maximum, and \textit{White Pulse} - a broad Gaussian distribution around both, represents white light. We compute the resulting exciton population weighted by oscillator strength:
\begin{equation} \begin{split}
    \Gamma&_{EEA}(T) = \\
    & \sum_{S_1,S_2} \bigg( \prod_{i=1,2} p(\omega-\Omega_{S_i,0}) M(S_i) \bigg) {\tau^{S_1,S_2}_{phEEA}}^{-1}(T)
\end{split} \end{equation}
Where $p$ is the Gaussian centered around $\omega$ and $M$ is the normalized oscillator strength, as calculated from GW-BSE.
Figure~\ref{ph_results}(D) shows the resulting annihilation rates. For all pulses, the rate varies slowly at low temperatures, while at higher temperatures its change is more rapid, with larger variations observed in Pulse $A$. This pulse contains both intra- and inter-valley coupling terms between $A_B-A_B$, $A_B-A_B'$, and its shape results directly from their increase at higher temperatures upon scattering of states with various momenta. Pulse $B$ results from the coupling terms $B-B$ and $B-B'$. These rates reflect the intrinsic phonon-assisted exciton–exciton scattering rates for the optically excited population under low-density conditions. 
While the overall magnitude and temperature dependence are consistent with expectations for phonon-activated annihilation processes~\cite{erkensten2021dark}, we emphasize that our results do not account for the full time evolution of the exciton population, which is expected to be further influenced by nonlinear effects arising at higher densities, as well as the above-discussed occupation of dark states in the ultra-fast timescales after excitation. 

In conclusion, we present a general ab initio framework to compute exciton-exciton annihilation in crystalline materials, incorporating both Coulomb- and phonon-mediated processes. Applying this method to monolayer WSe$_2$, we find that EEA occurs on timescales of hundreds of picoseconds and predominantly leads to the generation of unbound electron-hole pairs. Additionally, the phonon-assisted process is shown to activate inter-valley, dark-dark, and dark-bright annihilation channels, which can be identified in variations of the rate as a function of temperature. Our results connect crystal structure to exciton annihilation through bandstructure information, offering new insights into exciton relaxation pathways in low-dimensional materials and highlighting the role of phonons in non-radiative exciton recombination processes.

We thank Mauro Del Ben for the insightful discussions. Computational resources were provided by the ChemFarm local cluster at the Weizmann Institute of Science. This study was supported by a Minerva Foundation Grant (No. 7135421) and a European Research Council (ERC) Starting Grant (No. 101041159).

\clearpage
\twocolumngrid
\onecolumngrid

\begin{center}
  \textbf{Supplemental Material: Exciton-Exciton Annihilation Mediated by Many-Body Coulomb and Phonon Interactions: An Ab Initio Study}\\[1.5ex]
  Guy Vosco$^1$ and Sivan Refaely-Abramson$^1$\\
  [1.0ex]
  $^1$ Department of Molecular Chemistry and Materials Science, \\Weizmann Institute of Science, Rehovot 7610001, Israel
\end{center}
\vspace{1cm}

\noindent\textbf{I. Computational details}\\

The atomic structure and the wavefunctions and electronic bandstructure were computed using density-functional theory (DFT) within the Quantum Espresso code \cite{giannozzi2017advanced,giannozzi2009quantum,giannozzi2020quantum} with the Perdew-Burke-Ernzerhof (PBE) exchange-correlation functional \cite{perdew1996generalized}. We use a Bloch plane-wave basis-set, with fully relativistic norm-conserving pseudopotentials from the pseudo-dojo library \cite{van2018pseudodojo}, including spin-orbit coupling and spinor wavefunctions. We first relaxed the WSe$_2$ with 350 Ry wavefunction cutoff, resulting in a lattice parameter of 3.32 $\AA$. We then continue to compute the self-consistent electron density by employing a 36 × 36 × 1 uniform k-grid with a 90 Ry wavefunction cutoff energy.

Exciton states and the optical absorption spectrum are calculated within many-body perturbation theory within the GW-BSE approximation using the BerkeleyGW code \cite{deslippe2012berkeleygw}. The quasi-particle energy corrections were computed within the G$_0$W$_0$ approximation and the generalized plasmon-pole approximation \cite{hybertsen1986electron}. We used a 6 x 6 x 1 k-point grid with 6000 spinor bands and a 25 Ry cutoff of the screening function. We then expand our sampling through the Nonuniform Neck Subsampling (NNS) scheme \cite{da2017nonuniform} to capture the behavior of the electron wavefunctions accurately throughout the Brillouin zone. The optical properties were computed by solving the Bethe-Salpeter equation (BSE) \cite{rohlfing2000electron}. The excitonic energies were calculated by computing the electron-hole interaction kernel including 6 valence and 8 conduction bands and with a dielectric matrix that was calculated using 2000 bands on a 24 × 24 × 1 k-point grid with a 10 Ry screening cutoff. The interaction kernel was then interpolated onto a 72 × 72 × 1 grid with 4 valence and 6 conduction bands. The particle-particle interaction kernel which is used to calculate the excitonic wavefunctions and the exciton-exciton annihilation matrix elements computed to include 14 bands (6 valence and 8 conduction bands) with the a dielectric matrix that was calculated using 2000 bands and 10 Ry screening cutoff over a 12 x 12 x 1 grid, using the extended kernel option in BerkeleyGW code. These interaction kernels were then interpolated onto a 36 x 36 x 1 grid to evaluate the BSE Hamiltonians within the Tamm-Dancoff approximation for all the center-of-mass momentum excitons and exciton-exciton annihilation matrix elements using our in-house code based on the original BerkeleyGW code. The resulting BSE eigenvalues then scissor-shifted to match the excitonic energy computed on the 72 x 72 x 1 grid.

We compute the phononic bandstructure as well as electron-phonon coupling matrix elements using density functional perturbation theory \cite{giustino2017electron} within the Quantum Espresso and the EPW packages \cite{ponce2016epw}, on a 36 × 36 × 1 uniform q-grid with a convergence energy threshold of 10$^-18$ Ry. 

\vspace{1.0cm}
\noindent\textbf{II. Exciton-exciton annihilation coupling matrix elements}\\

The eight interaction channels presented in Figure 1 in the main text can be written in second-quantization formalism to establish the sign conventions from normal ordering and clarify momentum dependencies in the computed BerkeleyGW interaction terms:

\begin{equation} \begin{split}
    V^d_{Xe} & (S_1,Q_1;S_2,Q_2;\tilde{v},\tilde{c},\tilde{k}) = \\
    & \sum_{v_1,c_1,k_1} \sum_{v_2,c_2,k_2} A^{S_1,Q_1}_{v_1,c_1,k_1} \cdot A^{S_2,Q_2}_{v_2,c_2,k_2} \sum_{\alpha,\beta,\gamma,\delta} \bra{0}{\wick{
    \c1{\psi_{\tilde{c'}}}
    \c2{\psi^{\dagger}_{\tilde{v'}}}
    \hat{W}_{\alpha,\beta,\gamma,\delta}
    \c3{c^{\dagger}_{\alpha}}
    \c4{c^{\dagger}_{\beta}}
    \c5{c_{\gamma}}
    \c1{c_{\delta}}
    \c3{\psi^{\dagger}_{c_1}}
    \c5{\psi_{v_1}}
    \c4{\psi^{\dagger}_{c_2}}
    \c2{\psi_{v_2}}
    }}\ket{0}
    \\ & = - \sum_{v_1,c_1,k_1} \sum_{v_2,c_2,k_2} A^{S_1,Q_1}_{v_1,c_1,k_1} \cdot A^{S_2,Q_2}_{v_2,c_2,k_2} \cdot K^d(\tilde{k}-k_1,k_1,k_1+Q_1,c_1,v_1,\tilde{c},c_2) \cdot \delta_{v_2,\tilde{v}} \delta_{k_2,\tilde{k}+Q_1}
    % \\ & = - \sum_{v_1,c_1,k_1} \sum_{v_2,c_2,k_2} A^{S_1,Q_1}_{v_1,c_1,k_1} \cdot A^{S_2,Q_2}_{v_2,c_2,k_2} \cdot K^d(k_1-\tilde{k},k_2,\tilde{k},c_2,\tilde{c},v_1,c_1) \cdot \delta_{v_2,\tilde{v}} \delta_{k_2+Q_2,\tilde{k}+\tilde{Q}}
    \\ & = - \sum_{v_1,c_1,c_2,k_1} A^{S_1,Q_1}_{v_1,c_1,k_1} \cdot A^{S_2,Q_2}_{\tilde{v},c_2,\tilde{k}+Q_1} \cdot K^d(\tilde{k}-k_1,k_1,k_1+Q_1,c_1,v_1,\tilde{c},c_2)
\end{split} \end{equation}

\begin{equation} \begin{split}
    V^x_{Xe} & (S_1,Q_1;S_2,Q_2;\tilde{v},\tilde{c},\tilde{k}) = \\ 
    & \sum_{v_1,c_1,k_1} \sum_{v_2,c_2,k_2} A^{S_1,Q_1}_{v_1,c_1,k_1} \cdot A^{S_2,Q_2}_{v_2,c_2,k_2} \sum_{\alpha,\beta,\gamma,\delta} \bra{0}{\wick{
    \c1{\psi_{\tilde{c'}}}
    \c2{\psi^{\dagger}_{\tilde{v'}}}
    \hat{W}_{\alpha,\beta,\gamma,\delta}
    \c3{c^{\dagger}_{\alpha}}
    \c4{c^{\dagger}_{\beta}}
    \c1{c_{\gamma}}
    \c5{c_{\delta}}
    \c3{\psi^{\dagger}_{c_1}}
    \c5{\psi_{v_1}}
    \c4{\psi^{\dagger}_{c_2}}
    \c2{\psi_{v_2}}
    }}\ket{0}
    \\ & = + \sum_{v_1,c_1,k_1} \sum_{v_2,c_2,k_2} A^{S_1,Q_1}_{v_1,c_1,k_1} \cdot A^{S_2,Q_2}_{v_2,c_2,k_2} \cdot K^d(Q_1,k_1,\tilde{k},c_1,\tilde{c},v_1,c_2) \cdot \delta_{v_2,\tilde{v}} \delta_{k_2,\tilde{k}+Q_1}
    % \\ & = + \sum_{v_1,c_1,k_1} \sum_{v_2,c_2,k_2} A^{S_1,Q_1}_{v_1,c_1,k_1} \cdot A^{S_2,Q_2}_{v_2,c_2,k_2} \cdot K^d(-Q_1,k_2,k_1+Q_1,c_2,v_1,\tilde{c},c_1) \cdot \delta_{v_2,\tilde{v}} \delta_{k_2+Q_2,\tilde{k}+\tilde{Q}}
    \\ & = + \sum_{v_1,c_1,c_2,k_1} A^{S_1,Q_1}_{v_1,c_1,k_1} \cdot A^{S_2,Q_2}_{\tilde{v},c_2,\tilde{k}+Q_1} \cdot K^d(Q_1,k_1,\tilde{k},c_1,\tilde{c},v_1,c_2)
\end{split} \end{equation}

\begin{equation} \begin{split}
    V^d_{Xh} & (S_1,Q_1;S_2,Q_2;\tilde{v},\tilde{c},\tilde{k}) = \\
    & \sum_{v_1,c_1,k_1} \sum_{v_2,c_2,k_2} A^{S_1,Q_1}_{v_1,c_1,k_1} \cdot A^{S_2,Q_2}_{v_2,c_2,k_2} \sum_{\alpha,\beta,\gamma,\delta} \bra{0}{\wick{
    \c1{\psi_{\tilde{c'}}}
    \c2{\psi^{\dagger}_{\tilde{v'}}}
    \hat{W}_{\alpha,\beta,\gamma,\delta}
    \c3{c^{\dagger}_{\alpha}}
    \c2{c^{\dagger}_{\beta}}
    \c4{c_{\gamma}}
    \c5{c_{\delta}}
    \c1{\psi^{\dagger}_{c_1}}
    \c5{\psi_{v_1}}
    \c3{\psi^{\dagger}_{c_2}}
    \c4{\psi_{v_2}}
    }}\ket{0}
    \\ & = + \sum_{v_1,c_1,k_1} \sum_{v_2,c_2,k_2} A^{S_1,Q_1}_{v_1,c_1,k_1} \cdot A^{S_2,Q_2}_{v_2,c_2,k_2} \cdot K^d(k_2-k_1-Q_1,\tilde{k}+\tilde{Q},k_1+Q_1,\tilde{v},v_1,v_2,c_2) \cdot \delta_{c_1,\tilde{c}} \delta_{k_1,\tilde{k}}
    % \\ & = + \sum_{v_1,c_1,k_1} \sum_{v_2,c_2,k_2} A^{S_1,Q_1}_{v_1,c_1,k_1} \cdot A^{S_2,Q_2}_{v_2,c_2,k_2} \cdot K^d(k_1+Q_1-k_2,k_2,k_2+Q_2,c_2,v_2,v_1,\tilde{v}) \cdot \delta_{c_1,\tilde{c}} \delta_{k_1,\tilde{k}}
    \\ & = + \sum_{v_1, v_2,c_2,k_2} A^{S_1,Q_1}_{v_1,\tilde{c},\tilde{k}} \cdot A^{S_2,Q_2}_{v_2,c_2,k_2} \cdot K^d(k_2-\tilde{k}-Q_1,\tilde{k}+Q_1+Q_2,\tilde{k}+Q_1,\tilde{v},v_1,v_2,c_2)
\end{split} \end{equation}

\begin{equation} \begin{split}
    V^x_{Xh} & (S_1,Q_1;S_2,Q_2;\tilde{v},\tilde{c},\tilde{k}) = \\
    & \sum_{v_1,c_1,k_1} \sum_{v_2,c_2,k_2} A^{S_1,Q_1}_{v_1,c_1,k_1} \cdot A^{S_2,Q_2}_{v_2,c_2,k_2} \sum_{\alpha,\beta,\gamma,\delta} \bra{0}{\wick{
    \c1{\psi_{\tilde{c'}}}
    \c2{\psi^{\dagger}_{\tilde{v'}}}
    \hat{W}_{\alpha,\beta,\gamma,\delta}
    \c3{c^{\dagger}_{\alpha}}
    \c2{c^{\dagger}_{\beta}}
    \c4{c_{\gamma}}
    \c5{c_{\delta}}
    \c1{\psi^{\dagger}_{c_1}}
    \c4{\psi_{v_1}}
    \c3{\psi^{\dagger}_{c_2}}
    \c5{\psi_{v_2}}
    }}\ket{0}
    \\ & = - \sum_{v_1,c_1,k_1} \sum_{v_2,c_2,k_2} A^{S_1,Q_1}_{v_1,c_1,k_1} \cdot A^{S_2,Q_2}_{v_2,c_2,k_2} \cdot K^d(Q_2,k_2,k_1+Q_1,c_2,v_1,v_2,\tilde{v}) \cdot \delta_{c_1,\tilde{c}} \delta_{k_1,\tilde{k}}
    % \\ & = - \sum_{v_1,c_1,k_1} \sum_{v_2,c_2,k_2} A^{S_1,Q_1}_{v_1,c_1,k_1} \cdot A^{S_2,Q_2}_{v_2,c_2,k_2} \cdot K^d(-Q_2,\tilde{k}+\tilde{Q},k_2+Q_2,\tilde{v},v_2,v_1,c_2) \cdot \delta_{c_1,\tilde{c}} \delta_{k_1,\tilde{k}}
    \\ & = - \sum_{v_1,v_2,c_2,k_2} A^{S_1,Q_1}_{v_1,\tilde{c},\tilde{k}} \cdot A^{S_2,Q_2}_{v_2,c_2,k_2} \cdot K^d(Q_2,k_2,\tilde{k}+Q_1,c_2,v_1,v_2,\tilde{v})
\end{split} \end{equation}

\begin{equation} \begin{split}
    V_{EEA}^{S_1;S_2}(v,c,k) = & + V^d_{Xe}(S_1,Q_1,S_2,Q_2,v,c,k) + V^d_{Xe}(S_2,Q_2,S_1,Q_1,v,c,k) \\
    & + V^x_{Xe}(S_1,Q_1,S_2,Q_2,v,c,k) + V^x_{Xe}(S_2,Q_2,S_1,Q_1,v,c,k) \\
    & + V^d_{Xh}(S_1,Q_1,S_2,Q_2,v,c,k) + V^d_{Xh}(S_2,Q_2,S_1,Q_1,v,c,k) \\
    & + V^x_{Xh}(S_1,Q_1,S_2,Q_2,v,c,k) + V^x_{Xh}(S_2,Q_2,S_1,Q_1,v,c,k)
\end{split} \end{equation}

Where $\psi_n$ are electronic states, $c^\dagger$ and $c$ are the creation and annihilation operators, and $K^d$ is the direct interaction kernel which in the extended kernel formalism corresponds to the screened Coulomb interaction between two particles.

\begin{equation} \begin{split}
    K^d(Q,k',k,m',m,n',n) & = -\int{d^3rd^3r' \psi_{m',k'}(r) \cdot \psi_{m,k}^*(r) \cdot W(r,r') \cdot \psi_{n,k + Q}(r') \cdot \psi_{n',k' + Q}^*(r')}
\end{split} \end{equation}

\vspace{1.0cm}
\noindent\textbf{III. Broadening parameter of the energy conservation}\\

Our computations involve a single-parameter Gaussian broadening to the energy conservation conditions. The broadened density of states is given by:

\begin{equation}
    \rho(\Delta E,\sigma) = \frac{1}{\sigma\sqrt{2\pi}} \exp(-\frac{x^2}{2\sigma^2})
\end{equation}

\begin{figure}
\includegraphics[width=1.0\linewidth]{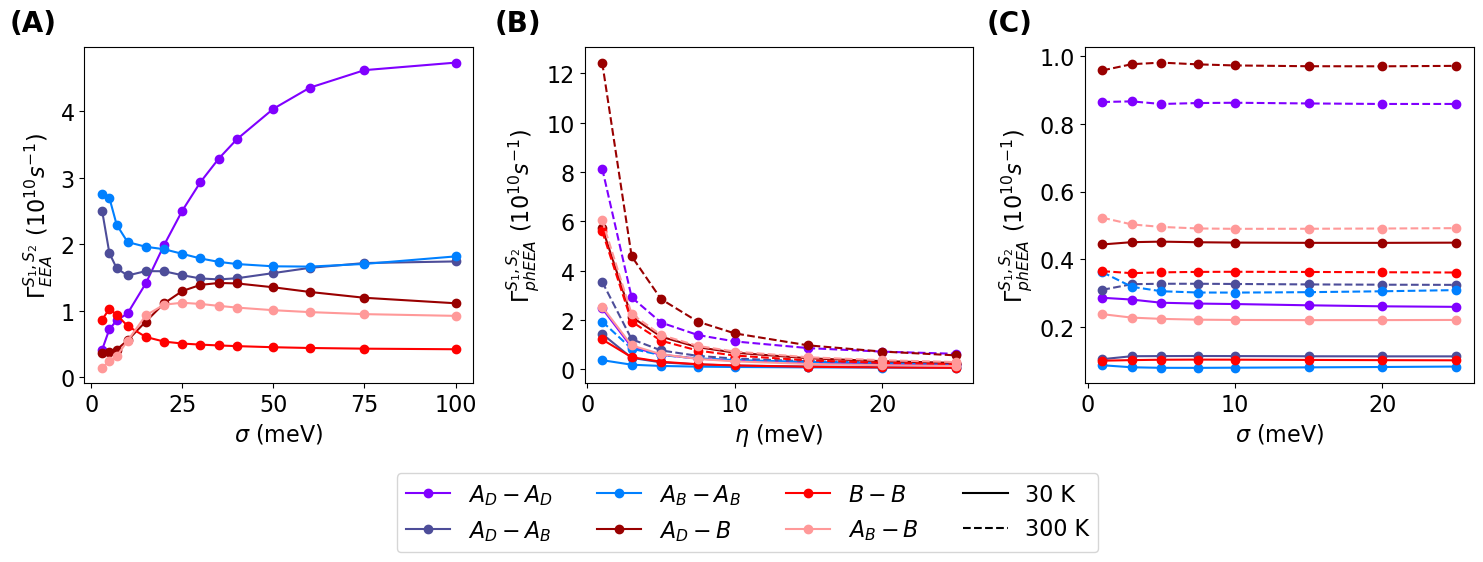}
\caption{(A) The Coulomb-driven EEA rates as a function of the energy conservation broadening. The phonon-assisted EEA rates (B) as a function of the intermediate state energy denominator parameter and (C) the initial-final energy conservation broadening at 30 K (solid lines) and 300 K (dashed lines).}
\label{dE_conv}
\end{figure}

This broadening accounts for numerical discretization and simulates physically allowed transitions between states that may not be explicitly sampled in our finite grids. We carefully examined the effect of the broadening parameter on our results, as presented in figure \ref{dE_conv} (A), to ensure that the results are not artificially influenced by summation over disallowed transitions. As expected, the transitions are unstable for small broadening; all of them stabilize around 40 meV up to a reasonable error, except for the $A_D-A_D$ transition, which only stabilizes at higher broadening. Still, the error from the broadening of 40 meV to the stabilization point is reasonable for this study.

Following the above, in the results presented in the manuscript, we used a 40 meV broadening parameter for the Coulomb-driven EEA. For the phonon-assisted EEA, the imaginary part of the self-energy appearing in the intermediate-state denominator requires a smaller broadening due to the narrower energy scales involved in the phonon dispersion. As shown in figure \ref{dE_conv}(B), the rates are sensitive to this intermediate-state broadening. In contrast, the energy conservation broadening has a smaller effect on the overall rate (panel C), suggesting that the denominator dominates the temperature and energy dependence. In the main text, we used a 10 meV broadening for energy conservation and a 10–20 meV range for the intermediate-state damping parameter.

\vspace{1.0cm}
\noindent\textbf{IV. Final states of the EEA}\\

We investigated the coupled final states allowing the EEA process, focusing on two regimes: the interactions of the dark and bright A peak excitons with A peak excitons, marked as A-A, and with B peak excitons, marked as A-B, which are coupled rather strongly as shown in the results presented in the main text. To reach a better understanding of the coupling to each examined final state, we sum over all the non-examined properties to integrate them out, showing the most probable band or momenta of the final states. The results are normalized for a trackable view of the most significant final states and hence do not represent the coupling magnitude. The final electron and hole bands, which undergo the strongest coupling to each initial state, are shown in Figure 3 (A), and in (B) we show their respective momentum. In order to better understand the allowed EEA coupling, we also studied the contribution of the final-state bands and momentum to the EEA rates, presented in Figure 3(C-D).

\begin{figure}
\includegraphics[width=1.0\linewidth]{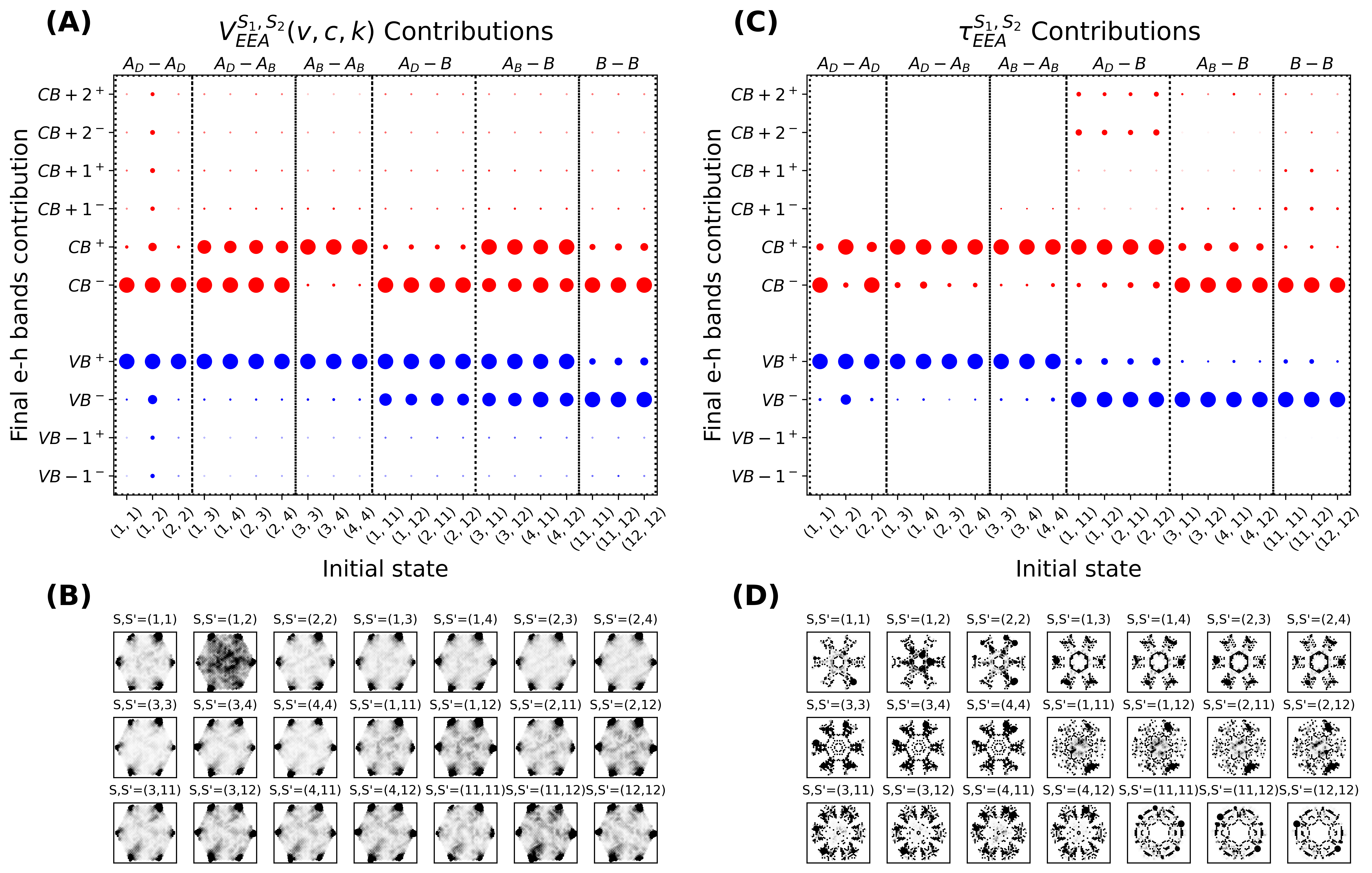}
\caption{Normalized final state (A) bands and (B) $k$-points contribution for EEA coupling matrix elements, (C-D) and to the EEA rate.}
\label{EEA_cont}
\end{figure}

The first ten columns in Figure \ref{EEA_cont}(A,C), along with the corresponding k-space maps in panels (B,D), represent the final electron and hole distributions resulting from the annihilation involving A-peak excitons, i.e. the A–A interactions. 
The first three columns correspond to annihilation between two dark excitons, marked as $A_D-A_D$. These combinations exhibit the strongest coupled electron–hole pairs in the lowest conduction band, $CB^-$, and highest valence bands, $VB^+$, near the $K$-valleys. However, since these transitions are forbidden by energy conservation at the K and K' points, the final electron–hole pair momenta shift toward the $\Lambda$ points, where energy and momentum conservation can be satisfied.
In contrast, when the two interacting excitons originate from different valleys, the coupling is very weak, and no well-defined final states are observed. This is because the destructive interference and weak wavefunction overlap between the initial excitons. The presented results are normalized, but the absolute coupling is significantly lower in these cross-valley cases.

The next four columns in Figure \ref{EEA_cont}(A,C) correspond to annihilation events between bright and dark A excitons, $A_D-A_B$. All four show similar coupling matrix element trends and produce final states located in the first two conduction bands and the top valence band, around the $K$-$\Lambda$ path.

The next three columns represent the annihilation of two bright excitons, $A_B-A_B$. These interactions yield final electron–hole pairs mainly at the second lowest conduction band, $CB^-$, and higher valance band, $VB^-$, now located along the path between $K$ and $M$.
Unlike the case of two dark excitons from different valleys, the annihilation between bright excitons does not lead to the cancellation of the coupling terms.
Although the coupling matrix elements are smaller than those for same-bright-exciton interactions, they remain significant.
The resulting final states show similar bands and momentum distributions due to the nearly equivalent excitons from the $K$ and $K'$ valleys.

The following eight columns and corresponding k-space maps represent the final electron–hole states resulting from annihilation events involving an A-peak and a B-peak exciton, i.e. the A–B interaction channels.
In the first four cases, where the interaction involves a dark A-peak exciton, $A_D-B$, the resulting hole mostly in the second highest valence band, $VB^-$, while the electron is in the second lowest conduction band, $CB^+$.
In contrast, the next four columns, which correspond to bright A-peak exciton interactions, $A_B-B$, yield final states where the electrons are mostly at the lowest conduction band, $CB^-$.

The last three columns correspond to annihilation between two B-peak excitons, $B-B$. These events produce electron–hole pairs located far from the band edge, in the second highest valance bands, $VB^-$, and lowest conduction band, $CB^-$.

In most cases, the dominance bands in the matrix elements also resulted in the final states while their momentum shifted to allow energy conservation. We note that although some contributions arise from the higher conduction bands $CB+2^\pm$, these are primarily associated with scattering involving higher-energy excitons (e.g., B-peak excitons 11–12) and dark states and are unlikely to play a significant role in EEA of A peak excitons.

\end{document}